\documentclass[showpacs,preprintnumbers,amsmath,amssymb]{revtex4}
\usepackage{graphicx}
\usepackage{dcolumn}
\usepackage{bm}%

\newtheorem{thm}{Theorem}
\newtheorem{cor}{Corollary}
\newtheorem{defi}{Definition}

\newtheorem{lem}{Lemma}
\newtheorem{rem}{Remark}

\newenvironment{proof}[1][Proof]{\textbf{#1.} }{\ \rule{0.5em}{0.5em}}

\newcommand{\C}{\mathbb C}

\newcommand{\R}{\mathbb R}

\def\la{\label}
\def\bt{\begin{thm}}
\def\et{\end{thm}}

\def\bl{\begin{lem}}
\def\el{\end{lem}}

\def\bd{\begin{defi}}
\def\ed{\end{defi}}

\def\bc{\begin{cor}}
\def\ec{\end{cor}}

\def\bp{\begin{proof}}
\def\ep{\end{proof}}

\def\br{\begin{rem}}
\def\er{\end{rem}}

\begin{document}

\title{Phase Transition and Separation for Mixture of Liquid  He-3 and He-4}

\thanks{The work was supported in part by the
Office of Naval Research and by the National Science Foundation.}
\author{Tian Ma}
\affiliation{Department of Mathematics, Sichuan University,
Chengdu, P. R. China}%

\author{Shouhong Wang}
 \homepage{http://www.indiana.edu/~fluid}
\affiliation{Department of Mathematics,
Indiana University, Bloomington, IN 47405}%
\date{\today}

\begin{abstract}
This article introduces a dynamical Ginzburg-Landau phase transition/separation model for the mixture of liquid helium-3 and helium-4, using a  unified dynamical Ginzburg-Landau model
for equilibrium phase transitions. 
The analysis of this model leads to three critical length scales $L_1 < L_2 < L_3$, detailed theoretical phase diagrams  and transition properties with different length scales of the container. 
\end{abstract}
\keywords{helium-3, helium-4, dynamic phase transition, lambda point, time-dependent Ginzburg-Landau models, dynamic transition theory}

\maketitle
\section{Introduction}
\label{sc1}


{\it Landau's legacy on  phase transition  has been a singular driving force for our recent work on developing  a general dynamic transition theory and a unified approach  for both equilibrium and non-equilibrium dynamic phase transitions. 
We hope that the work in this paper is in the spirit of Landau's legacy.
}

\bigskip

Superfluidity is a phase of matter  in which "unusual" effects are observed when liquids, typically of helium-4 or helium-3, overcome friction by surface interaction when at a stage, known as the "lambda point" for helium-4, at which the liquid's viscosity becomes zero.

 The main objectives of this article is to study $\lambda$-phase transitions of liquid helium-4 and and phase separations between liquid helium-3 and liquid helium-4 from both the  modeling and analysis points of view. 
 
In the late 1930s, Landau proposed a mean field theory of continuous phase transitions.  With the successful application of the Ginzburg-Landau theory to  superconductivity, it is nature to transfer something similar to the superfluidity case, as the superfluid transitions in liquid $^3$He  and  $^4$He   are of similar quantum origin as superconductivity. 
Unfortunately, we know that the classical Ginzburg-Landau free energy 
is poorly applicable to liquid helium in a quantitative sense, as described in by Ginzburg in \cite{ginzburg}.

As an attempt for this challenge problem, we introduces  a dynamical Ginzburg-Landau phase transition/separation model for the mixture of liquid helium-3 and helium-4. In this model, we use  an order parameter  $\psi$ for the phase transition of liquid $^4$He between the normal and superfluid states, 
and the mol fraction $u$ for liquid $^3$He. As $u$  is a conserved quantity, a Cahn-Hilliard type equation is needed for $u$, and a Ginzburg-Landau type equation is needed for the order parameter. The interactions of this quantities are built into the system naturally by using a  unified dynamical Ginzburg-Landau model
for equilibrium phase transitions, where the dynamic model is derived as a a gradient-type flow
as outlined in the appendix.

This analysis of the model established enables us to give a detailed study 
on the $\lambda$-phase transition and the phase separation between liquid $^3$He and $^4$He. In particular, we derived three critical length scales $L_1 < L_2 < L_3$ and the corresponding 
$\lambda$-transition and phase separation diagrams.  
The derive theoretical phase diagrams  based on our analysis agree with classical phase diagram, as shown e.g. in  Reichl \cite{reichl}  and  Onuki \cite{onuki}, and it is hoped that the study here will lead to a better understanding of mature of superfluids. 
Finally, we remark that  the order of second  transition is mathematically more challenging, and will be reported elsewhere. 

One important new ingredient for the analysis is  a new dynamic transition theory developed recently by the authors \cite{chinese-book, b-book}. With this theory, we derive a  new dynamic phase transition classification scheme, which   classifies phase transitions into three categories:  Type-I, Type-II and Type-III, corresponding respectively to the continuous, the jump and mixed transitions in the dynamic transition theory. 

\section{Model for Liquid Mixture of $^3$He -$^4$He}
Liquid $^3$He  and $^4$He  can be dissolved into each other. When $^3$He -atoms
are dissolved in liquid $^4$He and the density of $^3$He 
increases, the $\lambda$-transition temperature $T_{\lambda}$
decreases; see the liquid mixture phase diagram of $^3$He -$^4$He 
(Figure \ref{f8.46}),  where $X=n_3/(n_3+n_4), n_3$ and $n_4$ are the
atom numbers of $^3$He and $^4$He respectively. When $X=0,
T_{\lambda}=2.17K$,  where $\lambda$-phase transition takes
place and liquid $^4$He   undergoes a transition to superfluid phase from the
normal liquid phase. When $X=0.67$ and temperature decreases to
$T=0.87K$, i.e., at the triple point $C$ in Figure \ref{f8.46}, the liquid
mixture of $^3$He -$^4$He  has   a phase separation.
\begin{figure}[hbt]
  \centering
  \includegraphics[width=0.3\textwidth]{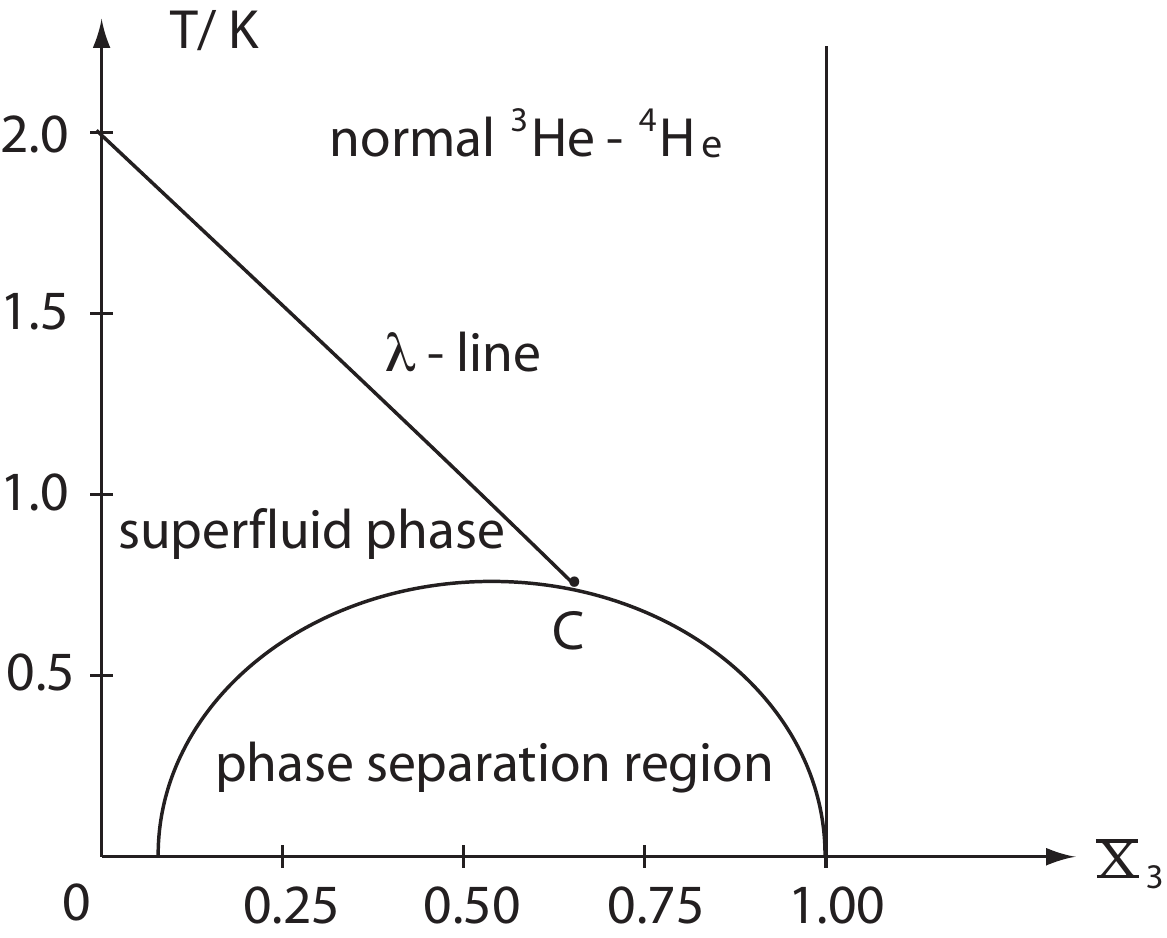}
  \caption{Liquid mixture phase diagram of $^3$He -$^4$He .}\la{f8.46}
 \end{figure}

Let the complex valued function $\psi=\psi_1 +  i \psi_2$ describe the superfluidity
of $^4$He,   and  $u$ be  the density of $^3$He, which is conserved, i.e., 
\begin{equation}
\int_{\Omega}udx=c\ \ \ \ (c>0\ \text{is\ a\ fixed\
number}).\label{8.257}
\end{equation}
The Ginzburg-Landau (Gibbs) free energy is
taken in the following form:
\begin{align}
G(\psi ,\rho_n,u)
=& \int_{\Omega}  \Big[  \frac{k_1 h^2}{2m} |\nabla \psi|^2 + \frac{\gamma_1}{2} |\psi|^2 
+\frac{\gamma_2}{4} |\psi|^4 
+ \frac{k_2}{2}   |\nabla u|^2   + \frac{\nu_1}{2} u^2
- \frac{\nu_2}{3} u^3+\frac{\nu_3}{4} u^4 + \nu_4 u |\psi|^2  \Big]dx. \label{8.258}
\end{align}
where $h$ is the Planck constant, $m$ is the mass of helium-4 atom,  and the coefficients satisfy 
\begin{equation}
\label{8.259}
k_1, k_2,  \gamma_2, \nu_2, \nu_3, \nu_4>0.
\end{equation}
When $\psi=0$, the liquid mixture of helium-3 and helium-4 is a
binary system, and $G(0, u)$ stands for the Cahn-Hilliard free energy.
By the standard model (\ref{7.38}), from (\ref{8.257}) and  (\ref{8.258}),  
the equations governing liquid mixture of $^3$He -$^4$He  are given as follows:
\begin{equation}
\begin{aligned} 
&\frac{\partial\psi}{\partial t}
= \frac{k_1h^2}{m} \Delta\psi - \gamma_1 \psi -\gamma_2 |\psi |^2\psi 
- 2 \nu_4 u\psi, \\
& 
\frac{\partial u}{\partial t}
= -k_2 \Delta^2u  + \Delta\big[ \nu_1 u -  \nu_2 u^2 + \nu_3 u^3 + \nu_4 |\psi|^2 \big].
\end{aligned}
\label{8.260}
\end{equation}
These equations have a physically sound constant steady state solution given by 
$$(\psi, u)=(0, u^0),$$
where $u^0 > 0$  is the density of helium-3 in a homogeneous state. 
For simplicity, we assume the total density $\rho=1$. Then the system control parameter $X$, the mol fraction of $^3$He, becomes 
$$X = u^0 \qquad \text{ with } 0 \le X \le 1.$$
Now we consider the derivations from this basic state:
$$(\psi, u) = (\psi', X + u'),$$
and we derive the following equations (drop
the primes):
 \begin{equation} 
 \begin{aligned}
 &\frac{\partial\psi}{\partial t}=\mu_1\Delta\psi  -  \lambda_1 \psi -\gamma_2|\psi |^2\psi
-2\nu_4 u\psi, \\
&\frac{\partial u}{\partial t}=-\mu_2 \Delta^2u  + \lambda_2\Delta
u+\Delta [  (- \nu_2 + 3 \nu_3 X) u^2+\nu_3u^3  + \nu_4 |\psi|^2 ], \\
& \int_{\Omega}udx=0,
\end{aligned}
\label{8.261}
\end{equation} 
with the Neumann boundary condition
\begin{equation}
\frac{\partial }{\partial n}
(u, \Delta u, \psi, \rho_n) =0 \qquad   \text{ on } \partial \Omega,
\label{8.262}
\end{equation}
where $mu_1=k_1h^2/m$, $\mu_2=k_2$, and 
\begin{equation}
\label{8.263}
\begin{aligned}
&  \lambda_1 = \gamma_1 + 2 \nu_4 X, \\
&     \lambda_2= \nu_1  -  2 \nu_2 X  + 3 \nu_3 X^2.
\end{aligned}
\end{equation}

\section{Phase Transitions of $^3$He -$^4$He  Mixtures}
We now  apply equations (\ref{8.260}) to study the phase transitions of $^3$He -$^4$He 
liquid mixtures. 

\subsection{Critical parameter curves}
For simplicity we only consider the special case
where  the container $\Omega$ is a rectangle:
$$\Omega =(0,L)\times (0,l)^2\subset \R^3 \qquad \text{ for }  L>l, $$
and the control parameters are the temperature $T$ and
the mol fraction $X$, and the length scale $L$  of the container $\Omega$.

Physically, by the Hildebrand theory (see \cite{reichl}),  in  the  lower temperature region and at $p=1$ atm, the critical parameter curve $\lambda_2 = 0$  is equivalent to 
\begin{equation}
T = \frac{2a}{R}(1-X)X -\sigma_0, 
\label{8.264}
\end{equation}
where $R$  is the molar gas constant and $a>0$ is a constant. Here $\sigma_0 >0$ is small correction term. The original Hildebrand theory leads to the case where $\sigma_0 =0$. However, as we can see from the classical phase separation of a binary system, the Hildebrand theory fails when the molar fraction is near $0$ or $1$, and the correction term added here agrees with the experimental  phase diagram as shown e.g. in Figure~4.13 in Reichl \cite{reichl}.

Furthermore, by (\ref{8.263}) and (\ref{8.264}), we have 
\begin{equation}
\label{8.264-1}
 \nu_1\simeq \theta_1 (T+ \sigma_0), \quad  3 \nu_3 \simeq 2 \nu_2, \quad  \frac{2 \nu_2}{\theta_1} \simeq \frac{2a}{R}.
\end{equation}

Consider eigenvalue problem of the linear operator in 
(\ref{8.261}): 
\begin{equation}
\left(
\begin{matrix}
\mu_1\Delta - \lambda_1 &0&0\\
0&\mu_1\Delta - \lambda_1&0\\
0 & 0&  -\Delta (\mu_2\Delta - \lambda_2) 
\end{matrix}
\right)
\left(\begin{matrix}
\psi_1\\
\psi_2\\
u
\end{matrix}
\right)  = \beta \left(\begin{matrix}
\psi_1\\
\psi_2\\
u
\end{matrix}
\right). \label{8.265}
\end{equation}
Here $\psi=\psi_1 + \psi_2$.
It is known that the first eigenvalue and  eigenvector of the
Laplacian operator with the Neumann condition and zero average 
 are $\lambda=\pi^2/L^2$ and $u=\cos\pi
x_1/L$.
Thus, the first two 
eigenvalues and their corresponding eigenvectors of (\ref{8.265}) are given  by 
\begin{align}
&  \left\{
\begin{aligned}
&  \beta_1
=- \lambda_1=-(\gamma_1 + 2 \nu_4 X),  \\
& (\psi_1,\psi_2,u)=(1,0,0)\ \text{and}\ (0,1,0),
\end{aligned}
\right.\label{8.266}\\
&
\left\{
\begin{aligned} 
& \beta_2 = -\frac{\pi^2}{L^2}\left(\frac{\mu_2\pi^2}{L^2}+ \lambda_2\right)
  =  -\frac{\pi^2}{L^2}
     \left(\frac{\mu_2\pi^2}{L^2}+\nu_1  -  2 \nu_2 X  + 3 \nu_3 X^2 \right), \\
&  (\psi_1,\psi_2,u) = (0, 0, \cos\pi x_1/L).
\end{aligned}
\right.\label{8.267}
\end{align}

As for $\nu_1$, the parameter $\gamma_1$ is approximately a linear function of $T$. Phenomenologically, we take 
\begin{equation}
 \gamma_1= -\sigma_1 + \theta_2 T\qquad (\theta_2, \sigma_1 > 0). \label{8.268}
\end{equation}
Then, by (\ref{8.264-1})  and (\ref{8.266})--(\ref{8.268}), the critical parameter curves in the $TX$-plane are as follows:
\begin{equation}
\label{8.269}
\begin{aligned}
& l_1: \beta_1=0  \qquad \Longleftrightarrow \qquad T_{c1}
  = \frac{\sigma_1 - 2 \nu_4 X}{\theta_2} = \frac{\sigma_1}{\theta_2} (1-X), \\
& l_2: \beta_2=0  \qquad \Longleftrightarrow \qquad T_{c2}=\frac{2 a }{R} X \left( 1-  X \right) -\sigma_0 - 
   \frac{\mu_2\pi^2}{\theta_1 L^2}.
   \end{aligned}
   \end{equation}
Here we have assumed that 
$\sigma_1 = 2 \nu_4.$
This assumption is true at $X=1$ as required by $T_{c1}=0$ at $X=1$, and is true as well 
at other values of $X$ as we  approximately take $\sigma_1$  and $\nu_4$ as constants. 

The critical parameter curve $l_1$ $(\beta_1=0)$ is as shown in
Figure \ref{f8.47}(a). 
Let 
\begin{equation}
L_0=\sqrt{\frac{\mu_1\pi^2}{\theta_1\left(\frac{a}{2R} - \sigma_0\right)}} \qquad \qquad \left(\sigma_0 <\frac{a}{2R}\right), \label{8.270}
\end{equation}
 then when $L>L_0$ the critical parameter curve
$l_2$  $(\beta_2=0)$ is as shown in Figure \ref{f8.47}(b).
\begin{figure}[hbt]
  \centering
  \includegraphics[width=0.2\textwidth]{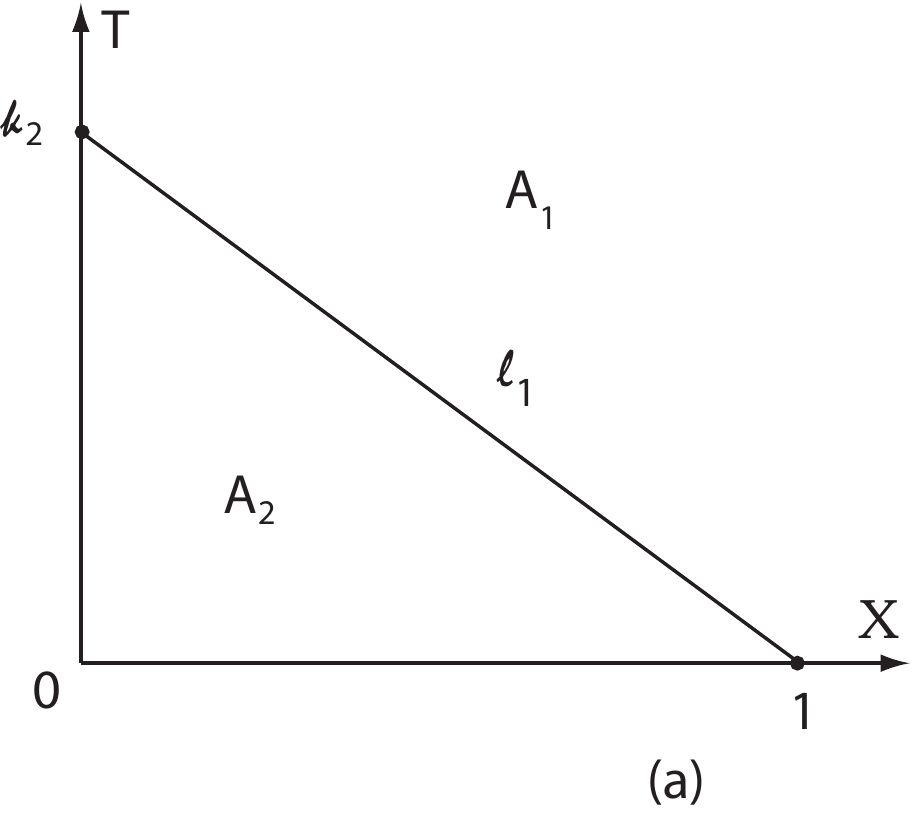}\qquad 
  \includegraphics[width=0.3\textwidth]{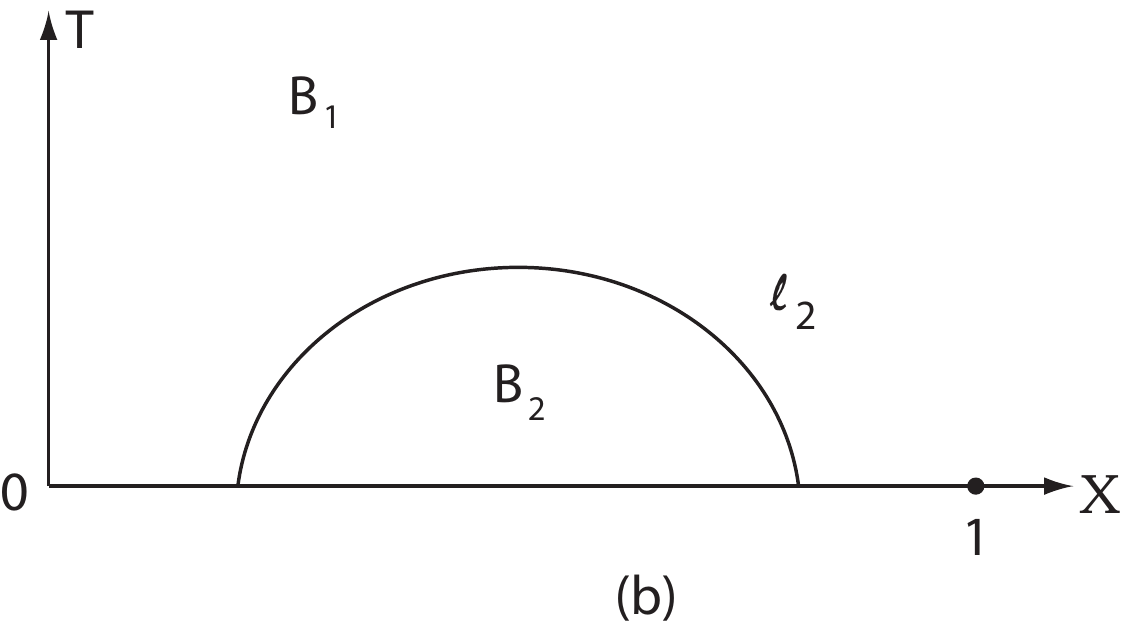}
  \caption{(a) $\beta_1 <0$ in region $A_1$, and
$\beta_1>0$ in $A_2$, (b) $\beta_2<0$ in region $B_1$, and
$\beta_2>0$ in $B_2$.}\la{f8.47}
 \end{figure}

\subsection{Transition theorems}

The following is the transition theorem for liquid mixture
$^3$He -$^4$He. Fro this purpose, we introduce a few length scales as follows:
\begin{align}
& L_1=\left(\frac{\mu_2\pi^2}{\theta_1}\right)^{1/2} \cdot 
\frac{1}{ \left( \frac{a}{2R} + \frac{2 \nu^2_4}{\theta_1 \gamma_2} - \sigma_0\right)^{1/2} }  \qquad \qquad \qquad \left(\frac{a}{2R} + \frac{2 \nu^2_4}{\theta_1 \gamma_2} - \sigma_0 > 0\right),  \label{8.270-1}\\
& \nonumber \\
&\label{8.270-2}
L_2 =\left\{
\begin{aligned}
& \left(\frac{\mu_2\pi^2}{\theta_1}\right)^{1/2}  \cdot 
   \frac{1}{  \left[  \frac{R}{8a} \left( \frac{2a}{R} - \frac{\sigma_1}{\theta_2}\right)^2 
+ \frac{2 \nu_4^2}{\theta_1 \gamma_2} - \sigma_0\right]^{1/2}}       && \text{ if }   \frac{R}{8a} \left( \frac{2a}{R} - \frac{\sigma_1}{\theta_2}\right)^2 
+ \frac{2 \nu_4^2}{\theta_1 \gamma_2} >  \sigma_0 \\
& \infty  && \text{ otherwise}, 
\end{aligned}  \right. \\
& \nonumber \\
&\label{8.270-3}
L_3 =\left\{
\begin{aligned}
& \left(\frac{\mu_2\pi^2}{\theta_1}\right)^{1/2}  \cdot 
   \frac{1}{  \left[  \frac{R}{8a} \left( \frac{2a}{R} - \frac{\sigma_1}{\theta_2}\right)^2 
 - \sigma_0\right]^{1/2}}       && \quad \qquad \text{ if }   \frac{R}{8a} \left( \frac{2a}{R} - \frac{\sigma_1}{\theta_2}\right)^2 
 >  \sigma_0 \\
& \infty  &&   \qquad \quad  \text{ otherwise}. 
\end{aligned}  \right. 
\end{align}
We remark here again that $\sigma_0$ is small as a correction term in the Hildebrand theory as mentioned earlier in (\ref{8.264}).

\bt\la{t8.19}  Let $T_{c1}$  and $T_{c2}$  be given in (\ref{8.269}). For equations (\ref{8.261}), we have the following assertions:

\begin{itemize}
\item[(1)] When
$L<L_1$ the system has only the superfluid phase
transition (i.e., the $\lambda$-phase transition) at $T=T_c^1$, and
the $TX$-phase diagram is as shown in Figure \ref{f8.47}(a).

\item[(2)]  Let $\sigma_1/\theta_2 < 4a/R$. Then $L_1 < L_2$, and 
if $L_1 < L < L_2$, then 
 there are two numbers  $0<X_1<\frac{1}{2} < X_2<1$ given by 
\begin{equation}
X_{1, 2} = \frac12 
\left[  1 \pm 
\sqrt{1 + \frac{2R}{a}\left( \frac{2 \nu_4^2}{\theta_1 \gamma_2}  -\sigma_0 
- \frac{\mu_2\pi^2}{\theta_1 L^2} \right) } \right] \label{8.270-3}
\end{equation}
such that the following hold true:

\begin{itemize}

\item[(a)] 
if $0\leq X<X_1$ or $X_2<X<1$,  the system has only the
$\lambda$-phase transition at $T=T_c^1$; 

\item[(b)] if  $X_1<X<X_2$, then  the system has
the $\lambda$-phase transition at $T=T_{c1}$, and has the phase
separation at $T^*=T_{c2} + \frac{2 \nu_4^2}{\theta_1 \gamma_2}< T_{c1}$. 
Moreover, the $TX$-phase diagram is as
shown in Figure \ref{f8.48}.
\end{itemize}

\item[(3)]  Let $\sigma_1/\theta_2 < 4a/R$. For any  $L_2 < L < L_3$, the phase diagram is as shown in Figure~\ref{f8.49-a}. For any $L_3 < L$, the phase diagram is as shown in Figure~\ref{f8.49-b}.

\item[(4)]  Let $\sigma_1/\theta_2 \ge 4a/R$. Then for any $L \ge L_1$, Assertion (2) holds true.

\item[(5)] The $\lambda$-phase transition at $T=T_c^1$ is  Type-I, which corresponds to second-order transition.
\end{itemize}
\et

\begin{figure}[hbt]
  \centering
  \includegraphics[width=0.4\textwidth]{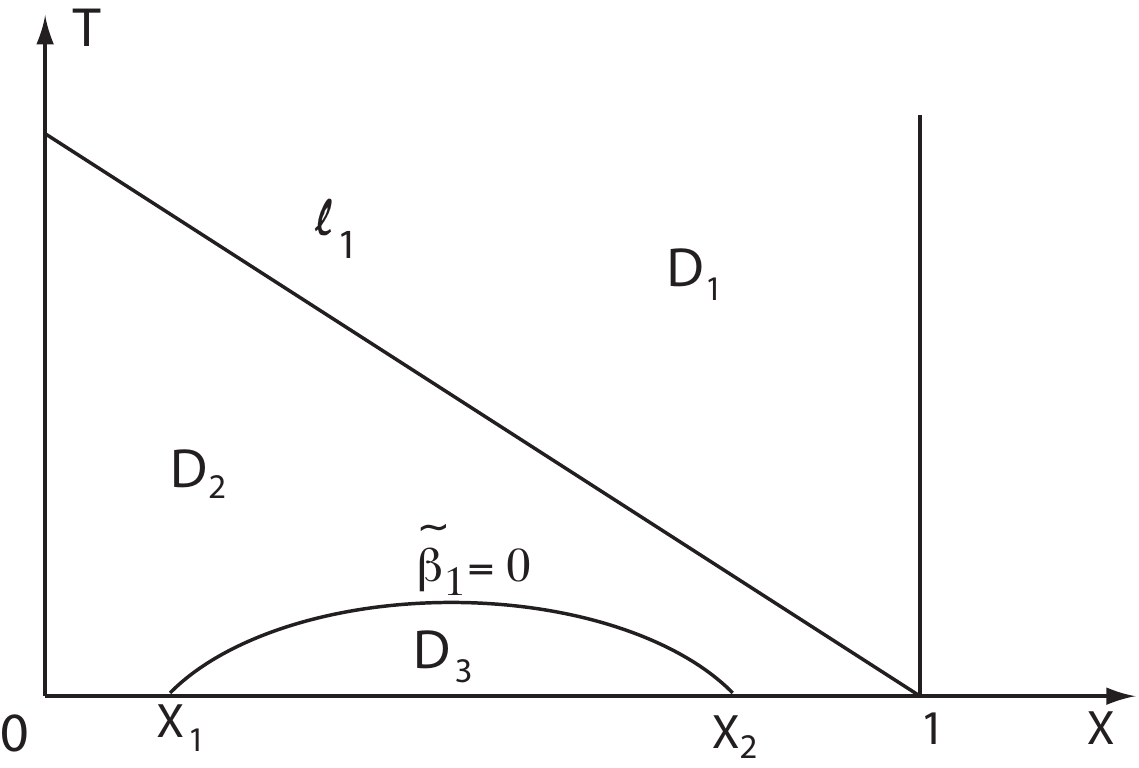}
  \caption{Region $D_1$ is normal $^3$He -$^4$He , $D_2$
is the superfluid phase, and $D_3$  is the phase separation region.}\la{f8.48}
 \end{figure}
\begin{figure}[hbt]
  \centering
  \includegraphics[width=0.4\textwidth]{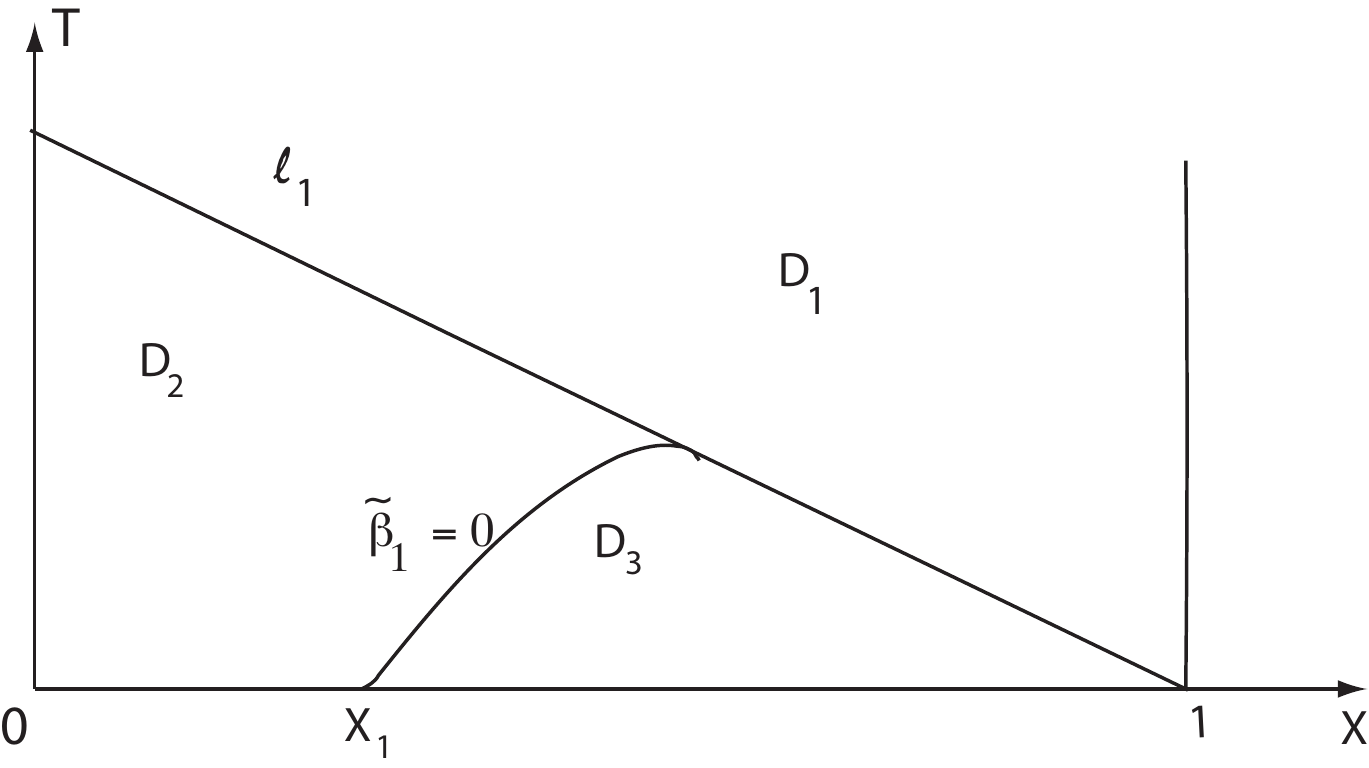}
  \caption{Region $D_1$ is normal $^3$He -$^4$He , $D_2$
is the superfluid phase, and $D_3$  is the phase separation region.}\la{f8.49-a}
 \end{figure}

\begin{figure}[hbt]
  \centering
  \includegraphics[width=0.4\textwidth]{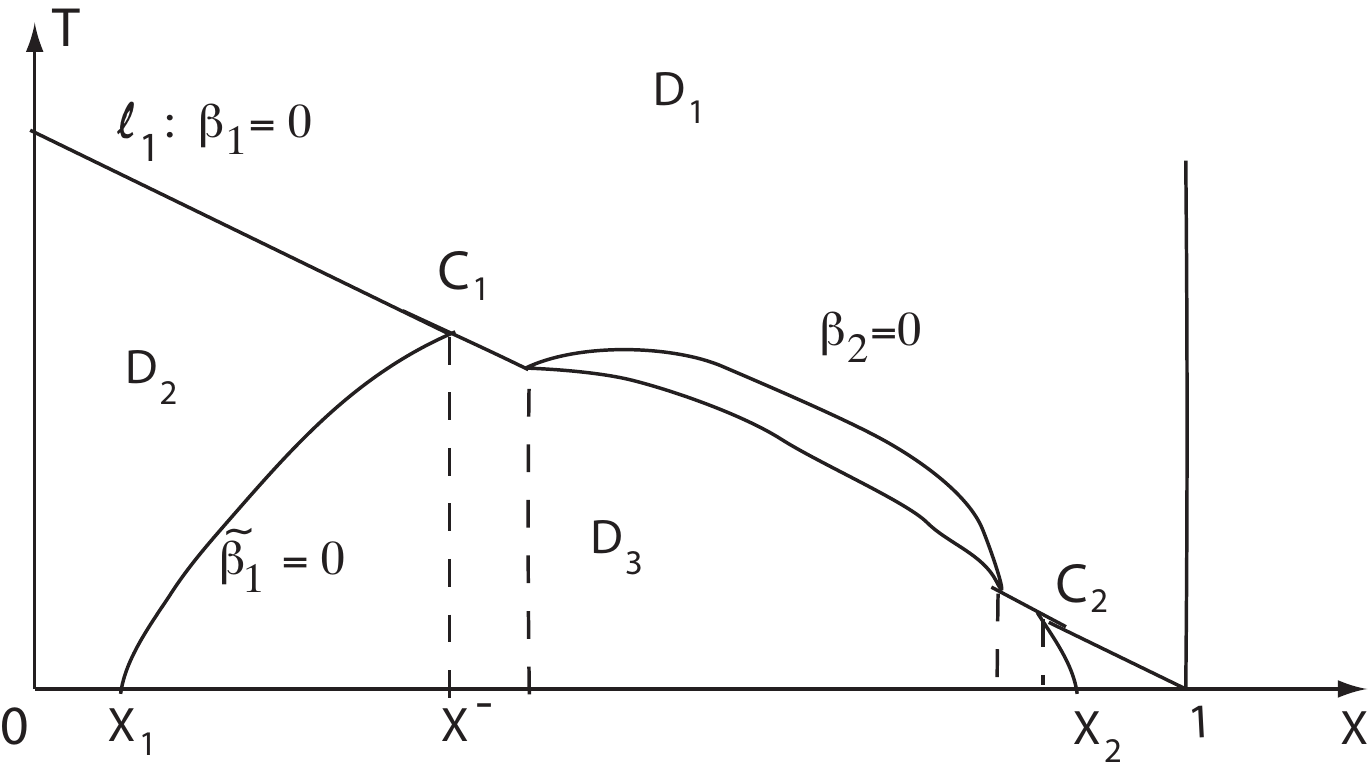}
  \caption{Region $D_1$ is normal $^3$He -$^4$He , $D_2$
is the superfluid phase, and $D_3$  is the phase separation region.}\la{f8.49-b}
 \end{figure}
 
 It is worth mentioning that the order of  second transitions crossing  $\widetilde{\beta}_1=0$ 
 is an interesting problem, which will be analyzed in a forthcoming paper.

\bp[Proof of Theorem \ref{t8.19}]
We proceed in several steps as follows.

\medskip

{\sc Step 1.} It is easy to see that the space
$$E=\{(u,\psi)=(0,\psi )|\ \psi\in \C\}$$
 is invariant for (\ref{8.261}).
Therefore, the transition solutions of (\ref{8.261})  from the critical parameter curve
$l_1$ $(\lambda_1=0)$ must be in $E$, which corresponds to the
superfluid transition. On the other hand, restricted on $E$,
equations (\ref{8.260}) are equivalent to the following ordinary differential equation:
\begin{equation}
\frac{d\psi}{dt}=- \lambda_1 \psi -\gamma_2 |\psi |^2\psi.
\label{8.271}
\end{equation}
It is then easy to see that  Assertion (5) holds true, and the bifurcated solutions consist of a circle given by 
$$\{ \sqrt{- \lambda_1/\gamma_2} \  e^{i\phi} \ | \  \phi \in [0, 2\pi]\}.$$

\medskip

{\sc Step 2.}  We now consider the second transition. For this purpose, 
let the transition solution of (\ref{8.260}) 
from $T=T_c^1$ (i.e., from $\lambda_1=0)$ be  given by
\begin{equation}
(0,\widetilde{\psi}(T))= (0,  \sqrt{- \lambda_1/\gamma_2} \  e^{i\phi})    \in E.\label{8.272}
\end{equation}
Take the transition
$$(u,\psi )\rightarrow (u^{\prime},
\psi^{\prime}+\widetilde{\psi}).
$$ 
Then the equations
(\ref{8.261}) are rewritten as (drop the primes)
\begin{equation} 
 \begin{aligned}
 &\frac{\partial\psi}{\partial t}=\mu_1\Delta\psi  +  \lambda_1  \psi   + \lambda_1 \psi^\ast-
 2 \nu_4 \sqrt{\frac{-\lambda_1}{\gamma_2}} u  
 - 2 \gamma_2 \sqrt{\frac{-\lambda_1}{\gamma_2}}  |\psi |^2
-2\nu_4 u\psi - \gamma_2|\psi |^2\psi, \\
&\frac{\partial u}{\partial t}=-\mu_2 \Delta^2u  + \lambda_2\Delta
u+ 2 \nu_4 \sqrt{\frac{-\lambda_1}{\gamma_2}} \Delta \psi_1
+ \Delta [  (- \nu_2 + 3 \nu_3 X) u^2+\nu_3u^3  + \nu_4 |\psi|^2 ], \\
& \int_{\Omega}udx=0,
\end{aligned}
\label{8.277}
\end{equation} 
where $\psi =\psi_1+i\psi_2$

We  consider the transition of
(\ref{8.277}) beyond $E$.
The linear operator of (\ref{8.277}) is given by 
\begin{equation}
B=\left(
\begin{matrix}
2 \lambda_1 & 0 &-2\nu_4 \sqrt{\frac{-\lambda_1}{\gamma_2}}  \\
& & \\
0 & \lambda_1  & 0\\
& & \\
2\nu_4 \sqrt{\frac{-\lambda_1}{\gamma_2}} \Delta & 0&  \Delta ( -\mu_2 \Delta + \lambda_2)
\end{matrix}
\right). \label{8.279}
\end{equation}
Restricted to 
its  first eigenspace 
$$
E_1=\left\{(\psi_1,\psi_2,u)=(y_1, y_2, y_3) \cos\frac{\pi x_1}{L}\ |\  (y_1, y_2, y_3) \in \R^3\right\}, 
$$
the linear $B$ is given by 
\begin{equation}
B|_{E_1}=\left(
\begin{matrix}
2 \lambda_1 & 0 &-2\nu_4 \sqrt{\frac{-\lambda_1}{\gamma_2}}  \\
& & \\
0 & \lambda_1 &  0\\
& & \\
- \frac{2\nu_4 \pi^2}{L^2} \sqrt{\frac{-\lambda_1}{\gamma_2}} & 0 & -\frac{\pi^2}{L^2} 
 ( \frac{\mu_2 \pi^2}{L^2}+ \lambda_2)
\end{matrix}
\right). \label{8.279}
\end{equation}
The eigenvalues $\widetilde{\beta}_1$, $\widetilde{\beta}_2$  and $\widetilde{\beta}_3$  of  $  B|_{E_1}$  are given by 
\begin{align*}
& \widetilde{\beta}_{1, 2} = \lambda_1
 - \frac{\pi^2}{2L^2}\left[ \frac{\mu_2\pi^2}{L^2}  + \lambda_2\right]
\pm  \sqrt{
\left[ \lambda_1 -\frac{\pi^2}{2L^2}  \left(\frac{\mu_2\pi^2}{L^2}  + \lambda_2\right) \right]^2 
+ \frac{2 \lambda_1 \pi^2}{L^2}
 \left( \frac{\mu_2\pi^2}{L^2}  + \lambda_2  - \frac{2 \nu_4^2}{\gamma_2}  \right)}, \\
& \widetilde{\beta}_3  =  \lambda_1 < 0. 
\end{align*}
We know that 
$$\widetilde{\beta}_3 < 0,  \quad \widetilde{\beta}_2 <  \widetilde{\beta}_1 \qquad \forall  T<  T_{c1}.
$$

By (\ref{8.264-1}), we have 
\begin{align*}
\lambda_2 &= \nu_1  -  2 \nu_2 X  + 3 \nu_3 X^2
=  \theta_1 (T+ \sigma_0) - 2 \nu_2(1-X)X \\
& =\theta_1\left(T+ \sigma_0- \frac{2\nu_2}{\theta_1}X(1-X)\right)\\
& = \theta_1 \left( T - T_{c2} \right) - \frac{\nu_2 \pi^2}{ L^2}
\end{align*}
Hence the transition curve $\widetilde{\beta}_1=0$ is given by 
$$ \frac{\mu_2\pi^2}{L^2}  + \lambda_2  = \frac{2 \nu_4^2}{\gamma_2}, $$
which is equivalent to 
$$  \theta_1 \left( T - T_{c2} \right) = \frac{2 \nu_4^2}{\gamma_2}. $$
Hence the transition curve $\widetilde{\beta}_1 =0$ is given by 
\begin{equation}\label{8.280}
\widetilde{\beta}_1 =0 \qquad \Longleftrightarrow \qquad T
= T_{c2} +\frac{2 \nu_4^2}{\theta_1\gamma_2}= \frac{2a}{R}X(1-X) -\sigma_0 - 
  \frac{\mu_2\pi^2}{\theta_1 L^2}  + \frac{2 \nu_4^2}{\theta_1 \gamma_2}.  
   \end{equation}
Then  $L_1$  defined by (\ref{8.270-1}) is the critical length scale to make 
the transition curve $\widetilde{\beta}_1 =0$ achieving  its maximum  at 
$(T, X)=(0, 1/2)$.

By definition, it is then easy to see that if $L< L_{1}$,  no phase separation occurs at any temperature although phase transition for $^4$He does occur as the temperature decreases below certain critical temperature.

\medskip

{\sc Step 3.} 
Now we calculate the length scale $L_2$ where the critical curves $\beta_1=0$  and $\widetilde{\beta}_1=0$ are tangent to each other, i.e., they interact at exactly one point.
By definition, we need to solve $L$ such that the equation
$$\frac{\sigma_1}{\theta_2} (1-X) = \frac{2 a }{R} X \left( 1-  X \right) -\sigma_0 - 
   \frac{\mu_2\pi^2}{\theta_1 L^2}$$
has exactly one solution.
Hence, it is easy to derive that the formula $L_2$   is as given by  (\ref{8.270-2}).

By comparing $L_1$  and $L_2$, we have 
\begin{equation}
 L_2 > L_1 \qquad \text{  if and only if } \qquad \frac{\sigma_1}{\theta_2} < \frac{4a}{R}.
 \end{equation}

Now we the length scale $L_3$ where the critical curves $\beta_1=0$  and $\beta_2=0$ are tangent to each other, i.e.,  we need to solve $L$ such that the equation
$$\frac{\sigma_1}{\theta_2} (1-X) = \frac{2 a }{R} X \left( 1-  X \right) -\sigma_0 $$
has exactly one solution. The formula $L_3$  is  as given by  (\ref{8.270-3}).

\medskip 

{\sc Step 4.} 
Now we return to prove Assertions (1)-(4). First, we consider the case where  
$\frac{\sigma_1}{\theta_2} < \frac{4a}{R}$, i.e., $L_2 > L_1$. 
In this case, for $L_1 < L < L_2$, second transition curve  $\widetilde{\beta}_1=0$ is as shown in Figure~\ref{f8.48}. Namely, as one decreases the temperature $T$ entering from region $D_2$ into $D_3$ through the curve $\widetilde{\beta}_1=0$, phase separation occurs. 

By  (\ref{8.280}), solving
$$\frac{2a}{R}X(1-X) -\sigma_0 - 
  \frac{\mu_2\pi^2}{\theta_1 L^2}  + \frac{2 \nu_4^2}{\theta_1 \gamma_2}=0$$
gives  the least and biggest mol fractions $X_1 < X_2$ defined by (\ref{8.270-3}), and  Assertion (2) follows. Other assertions can be proved in the same fashion.

The proof of the theorem is complete.
\ep

\section{Physical conclusions}
We have introduced a dynamical Ginzburg-Landau phase transition/separation model for the mixture of liquid helium-3 and helium-4. In this model, we use  an order parameter  $\psi$ for the phase transition of liquid $^4$He between the normal and superfluid states, 
and the mol fraction $u$ for liquid $^3$He. As $u$  is a conserved quantity, a Cahn-Hilliard type equation is needed for $u$, and a Ginzburg-Landau type equation is needed for the order parameter. The interactions of this quantities are built into the system naturally by using a  unified dynamical Ginzburg-Landau model
for equilibrium phase transitions, where the dynamic model is derived as a a gradient-type flow
as outlined in the appendix. 
 
This analysis of the model established enables us to give a detailed study 
on the $\lambda$-phase transition and the phase separation between liquid $^3$He and $^4$He. In particular, we derived three critical length scales $L_1 < L_2 < L_3$ with the following conclusions:
\begin{itemize}

\item[1)] For $L< L_1$, there is only $\lambda$-phase transition for $^4$He and no phase separation  between 
$^3$He and $^4$He,  as shown in Figure \ref{f8.47}(a).

\item[2)] For $L_1< L< L_2$, there is no triple points, and phase separation occurs as a second phase transition
after the $\lambda$-transition when the mol fraction is between two critical values, as shown in Figure \ref{f8.48}.

\item[3)] For $L_2 < L < L_3$, the $\lambda$-transition is always the first transition,   as shown in Figure~\ref{f8.49-a}.

\item[4)]  
For $L_3 < L$,  both the $\lambda$-transition and the phase separation can appear as either the first transition or the second transition depending on the mol fractions, as shown in Figure~\ref{f8.49-b}. In this case, when the phase separation is the first transition, the separation mechanism is the same as a typical binary system as described in great detail by the authors in 
\cite{MW08d}.

\end{itemize}
Also, the $\lambda$-transition is always second-order. 
The derive theoretical phase diagram  based on our analysis agrees with classical phase diagram, and it is hoped that the study here will lead to a better understanding of mature of superfluids. 
Finally, we remark that  the order of second  transition is mathematically more challenging, and will be reported elsewhere.

\appendix
\section{Dynamic Ginzburg-Landau models for equilibrium phase transitions}
\label{s7.2.2}
In this section, we recall a unified  time-dependent Ginzburg-Landau theory for modeling  equilibrium phase transitions in statistical physics; see also \cite{MW08f, MW08c}.

Consider a thermal system with a control  parameter $\lambda$. 
By the mathematical characterization of gradient systems and the le Ch\^atelier principle, for a system with
thermodynamic potential ${\mathcal{H}}(u,\lambda )$, the governing
equations are essentially determined by the functional
${\mathcal{H}}(u,\lambda )$.
When the order parameters $(u_1,\cdots,u_m)$ are nonconserved
variables, i.e., the integers
$$\int_{\Omega}u_i(x,t)dx=a_i(t)\neq\text{constant}.$$
then the time-dependent equations are given by
\begin{equation}
\left.
\begin{aligned} 
&\frac{\partial u_i}{\partial
t}=-\beta_i\frac{\delta}{\delta u_i}{\mathcal{H}}(u,\lambda
)+\Phi_i(u,\nabla u,\lambda ),
\end{aligned}
\right.\label{7.30}
\end{equation}
for  $1 \le i \le m$, where $\beta_i>0$ and $\Phi_i$ satisfy
\begin{equation}
\int_{\Omega}\sum_i\Phi_i\frac{\delta}{\delta
u_i}{\mathcal{H}}(u,\lambda )dx=0.\label{7.31}
\end{equation}
The condition (\ref{7.31})  is  required by
the Le Ch\^atelier principle. In the concrete problem, the terms
$\Phi_i$ can be determined by physical laws and (\ref{7.31}). We remark here that following the le Ch\^atelier principle, one should have an inequality constraint. However   physical systems often obey most simplified rules, as  many existing models for specific problems are consistent with the equality constraint here. This remark applies to the constraint (\ref{7.37}) below as well.

When the order parameters are the number density and the system
has no material exchange with the external, then $u_j$  $(1\leq j\leq
m)$ are conserved, i.e.,
\begin{equation}
\int_{\Omega}u_j(x,t)dx=\text{constant}.\label{7.32}
\end{equation}
This conservation law requires a continuity equation
\begin{equation}
\frac{\partial u_j}{\partial t}=-\nabla\cdot J_j(u,\lambda
),\label{7.33}
\end{equation}
where $J_j(u,\lambda )$ is the flux of component $u_j$, satisfying
\begin{equation}
J_j=-k_j\nabla (\mu_j-\sum_{i\neq j}\mu_i),\label{7.34}
\end{equation}
where $\mu_l$ is the chemical potential of component $u_l$, 
\begin{equation}
\mu_j-\sum_{i\neq j}\mu_i=\frac{\delta}{\delta
u_j}{\mathcal{H}}(u,\lambda )-\phi_j(u,\nabla u,\lambda
), \label{7.35}
\end{equation}
and  $\phi_j(u,\lambda )$ is a function depending on the other
components $u_i$ $(i\neq j)$. Thus, from
(\ref{7.33})-(\ref{7.35}) we obtain the dynamical equations as
follows
\begin{equation}
\begin{aligned} 
&\frac{\partial u_j}{\partial
t}=\beta_j\Delta\left[\frac{\delta}{\delta
u_j}{\mathcal{H}}(u,\lambda )-\phi_j(u,\nabla u,\lambda )\right],
\end{aligned}
\label{7.36}
\end{equation}
for $1 \le j \le m$, where $\beta_j>0$ are constants,  and  $\phi_j$ satisfy
\begin{equation}
\int_{\Omega}\sum_j\Delta\phi_j\cdot\frac{\delta}{\delta
u_j}{\mathcal{H}}(u,\lambda )dx=0.\label{7.37}
\end{equation}

When $m=1$, i.e., the system is a binary system, consisting of
two components $A$ and $B$, then the term $\phi_j=0$. The above model covers the classical Cahn-Hilliard model. It is worth mentioning that for multi-component systems, these $\phi_j$ play an important rule in deriving good time-dependent models.

If the order parameters $(u_1,\cdots,u_k)$ are coupled to the
conserved variables $(u_{k+1},\cdots,u_m)$, then the dynamical
equations are
\begin{equation}
\begin{aligned} 
&\frac{\partial u_i}{\partial t}
   =-\beta_i\frac{\delta}{\delta u_i}{\mathcal{H}}(u,\lambda)+\Phi_i(u,\nabla u,\lambda ),\\
& \frac{\partial u_j}{\partial t}
  =\beta_j\Delta\left[\frac{\delta}{\delta u_j}{\mathcal{H}}(u,\lambda )
    -\phi_j(u,\nabla u,\lambda )\right],\\
\end{aligned}
\label{7.38}
\end{equation}
for  $ 1 \le i \le k$  and $k+1 \le j \le m$.
Here $\Phi_i$  and $\phi_j$ satisfy (\ref{7.31})   and (\ref{7.37}), respectively.

The model (\ref{7.38}) we derive here  gives a general form of the governing
equations to thermodynamic phase transitions, and will play crucial role in studying the dynamics of equilibrium phase transitions in statistical physics.

\bibliographystyle{siam}
\def\cprime{$'$}

\end{document}